\documentclass[sigconf]{acmart}
\usepackage[flushleft]{threeparttable}
\usepackage{xspace}
\usepackage{marvosym}
\usepackage{multirow}
\usepackage[table,xcdraw]{xcolor}
\AtBeginDocument{%
  }

\setcopyright{acmlicensed}
\copyrightyear{2026}
\acmYear{2026}
 \acmDOI{XXXXXXX.XXXXXXX}
\acmConference[WWW'26]{the Web Conference 2026}{April 13--16, 2026}{Dubai, UAE}
\acmISBN{978-1-4503-XXXX-X/2018/06}

\def\bx{\mathbf{x}}
\def\by{\mathbf{y}}
\def\bz{\mathbf{z}}

\begin{document}

\title{Warmer for Less: A Cost-Efficient Strategy for Cold-Start Recommendations at Pinterest}


\author{Saeed Ebrahimi$^*$}
\email{mebrahimisaadabadi@pinterest.com}
\affiliation{%
  \institution{Pinterest}
  \city{San Francisco}
  \state{CA}
\country{USA}
}

\author{Weijie Jiang$^*$}
\email{weijiejiang@pinterest.com}
\affiliation{%
  \institution{Pinterest}
  \city{San Francisco}
  \state{CA}
\country{USA}
}

\author{Jaewon Yang$^\text{\Cross}$}
\email{jaewonyang@pinterest.com}
\affiliation{%
  \institution{Pinterest}
  \city{San Francisco}
  \state{CA}
  \country{USA}
}

\author{Olafur Gudmundsson}
\email{ogudmundsson@pinterest.com}
\affiliation{%
 \institution{Pinterest}
  \city{San Francisco}
  \state{CA}
 \country{USA}
 }

\author{Yucheng Tu}
\email{ytu@pinterest.com}
\affiliation{%
 \institution{Pinterest}
  \city{San Francisco}
  \state{CA}
 \country{USA}
 }

\author{Huizhong Duan}
  \email{hduan@pinterest.com}
\affiliation{%
  \institution{Pinterest}
  \city{San Francisco}
  \state{CA}
  \country{USA}
  }




\renewcommand{\shortauthors}{Trovato et al.}

\def\onedot{.\xspace}
\def\eg{\emph{e.g}\onedot} \def\Eg{\emph{E.g}\onedot}
\def\ie{\emph{i.e}\onedot} \def\Ie{\emph{I.e}\onedot}
\def\cf{\emph{cf}\onedot} \def\Cf{\emph{Cf}\onedot}
\def\etc{\emph{etc}\onedot} \def\vs{\emph{vs}\onedot}
\def\wrt{w.r.t\onedot} \def\dof{d.o.f\onedot}
\def\iid{i.i.d\onedot} \def\wolog{w.l.o.g\onedot}
\def\etal{\emph{et al}\onedot}

\begin{abstract}
Pinterest is a leading visual discovery platform where recommender systems (RecSys) are key to delivering relevant, engaging, and fresh content to our users. In this paper, we study the problem of improving RecSys model predictions for \emph{cold-start} (CS) items, which appear infrequently in the training data. Although this problem is well-studied in academia, few studies have addressed its root causes effectively at the scale of a platform like Pinterest.

By investigating live traffic data, we identified several challenges of the CS problem and developed a corresponding solution for each:
First, industrial-scale RecSys models must operate under tight computational constraints. Since CS items are a minority, any related improvements must be highly cost-efficient. To address this, our solutions were designed to be lightweight, collectively increasing the total parameters by only 5\%.
Second, CS items are represented only by non-historical (e.g., content or attribute) features, which models often treat as less important. To elevate their significance, we introduce a residual connection for the non-historical features.
Third, CS items tend to receive lower prediction scores compared to non-CS items, reducing their likelihood of being surfaced. We mitigate this by incorporating a score regularization term into the model.
Fourth, the labels associated with CS items are sparse, making it difficult for the model to learn from them. We apply the manifold mixup technique to address this data sparsity.
Implemented together, our methods increased fresh content engagement at Pinterest by 10\% without negatively impacting overall engagement and cost, and have been deployed to serve over 570 million users on Pinterest.
\end{abstract}

\vspace{-1em}


\received{20 February 2007}
\received[revised]{12 March 2009}
\received[accepted]{5 June 2009}

  \maketitle
\def\thefootnote{*}\footnotetext{Both the authors contributed equally to this work.}\def\thefootnote{\arabic{footnote}}
\def\thefootnote{\Cross}\footnotetext{Corresponding author}\def\thefootnote{\arabic{footnote}}


\section{Introduction}

\begin{figure*}
  \centering
  \includegraphics[width=0.9\linewidth]{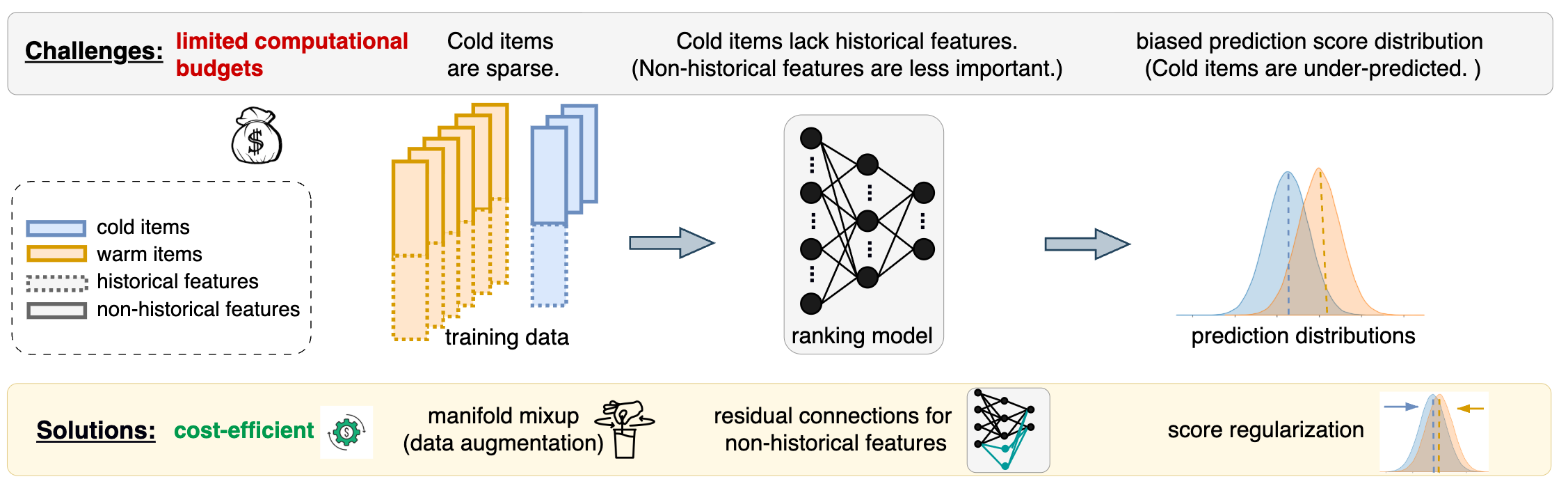}
  \vspace{-8pt}
  \Description{overview and comparison with other methods.}
  \caption{
  The overview of key challenges in CS recommendation and how the proposed framework addresses them.}
\end{figure*}

At Pinterest, a visual search and discovery platform with over 570 million users and 1.5 billion Pins (items representing ideas) saved each week, Recommender Systems (RecSys) are essential for delivering relevant and engaging Pins personalized to each user and context \cite{chang2023latent, chen2025pinfm, pancha2022pinnerformer}.
The main ranking model in RecSys is especially important since it predicts action probabilities for each candidate and ultimately determines which items are surfaced to users \cite{badrinath2025omnisage, xia2023transact, agarwal2024omnisearchsage}. In particular, generating unbiased predictions for fresh content is critical because it provides temporal relevance, promotes creator and merchant equity, sustains ecosystem health, and improves user experience \cite{chen2025cold, zhao2023bootstrapping}.


In this paper, we address the challenge of improving ranking model accuracy for fresh items that lack substantial engagement history, a phenomenon known as the cold-start (CS) problem \cite{jiang2024prompt, monteil2024marec, li2024scene}. Because ranking models are typically trained on historical user engagement data, they often develop biases towards well-established content that frequently appears in the training set \cite{zhang2022incorporating, joachims2017unbiased}. This serving bias can result in stale recommendations, repeatedly surfacing the same content \cite{chen2024multi, ganhor2024multimodal, chen2025cold, zhao2023bootstrapping}. Successfully identifying engaging CS Pins can break this feedback loop, introducing serendipity and bringing a refreshing experience to users.

Addressing the CS problem in ranking models at Pinterest’s scale presents several challenges. (1) Industrial recommender systems operate under strict computational budgets for both training and serving. Since CS items are, by definition, a minority, it is important to solve the CS problem without significantly increasing computational costs. (2) CS items often lack key features used by ranking models, especially those based on user engagement history \cite{pan2019warm, zhu2020recommendation}, known as historical features. (3) CS items tend to show lower observed engagement rates in training data, partly because existing production models seldom promote them \cite{huang2023aligning}. As a result, ranking models learn to underestimate the action probabilities for CS items. (4) CS items are infrequently shown to users, which further exacerbates data sparsity \cite{zhu2021learning}.

Research have proposed a variety of approaches towards these challenges. Cost-efficient methods such as dropout usually leads to tradeoff between CS and non-CS items \cite{volkovs2017dropoutnet, zhao2023bootstrapping}. Content-based methods mitigate the low coverage of historical features on CS items \cite{wei2021contrastive,barkan2019cb2cf}. However, their impact is limited due to the gap between the semantic information and the recommendation tasks \cite{jiang2024prompt, chen2023win}, \ie, users do not always engage with items that are semantically similar to each other or to their search queries. Knowledge distillation frameworks aim to mitigate model's underestimation on CS items \cite{zhu2020recommendation, wei2021contrastive}, however, distillation requires complicated tuning and significant increase of model parameters from the additional student tower, leading to computational overhead and extra infra cost. Transfer learning based methods deal with cold item sparsity issue, but they usually require another model (cost incurred) \cite{barjasteh2016cold, zhu2020recommendation}, such as clustering methods to group items and facilitate the learning from popular items to cold items \cite{chang2024cluster}. 


\begin{figure*}[t]
  \centering
  \includegraphics[width=1\linewidth]{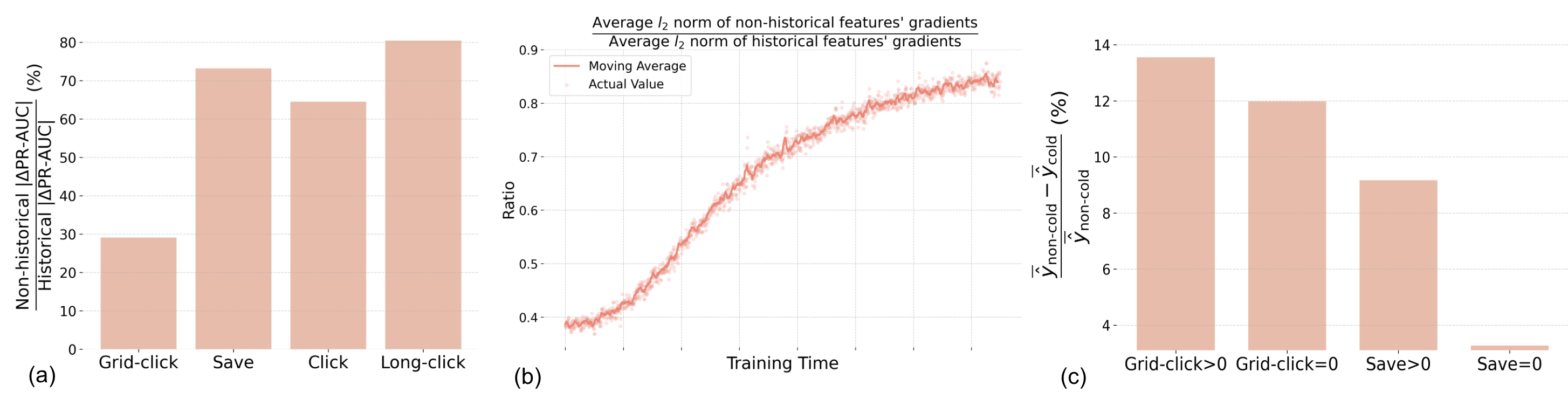}
  \caption{(a) $\Delta$PR-AUC ratio from ablating historical and non-historical features, indicating stronger dependence on historical signals, with the ratio staying well below 100\%. (b) Ratio of average $\ell_2$-norm of prediction gradients with respect to historical and non-historical features during training, illustrating that model updates are dominated by historical signals, as the ratio remains below one. (c) Visualization of predicted scores for randomly selected warm and cold samples (both positive and negative samples), illustrating that the model assigns substantially higher scores to warm items. This prediction bias leads to recommendations dominated by historical-rich items and systematic under-recommendation of cold items.}\label{motivation1}
  \Description{A figure showing the results of the feature importance and score bias experiment.}
  \label{motivation}
\end{figure*}

In this paper, we propose solutions to the CS problem for web-scale recommender systems. We systematically address the challenges outlined above, present a novel insight based on data analysis from Pinterest, implement a solution that maintains the existing training and serving costs of ranking models, and deploy it to production. To our knowledge, this is one of the few works to systematically analyze the root causes of the CS problem, propose solutions, and validate them through large-scale A/B testing.

Here are the solutions we present for each challenge.
For the first bottleneck---the computational constraint---we demonstrate that all proposed solutions are cost-effective.
For the second challenge---the lack of historical features for CS items---our analysis revealed that (1) historical features are significantly more important than non-historical features, and (2) during model training, ranking model relies less on non-historical features. In Figure~\ref{motivation1}(a), we show the ratio of the PR-AUC drop ($\Delta$PR-AUC) when either historical or non-historical features are removed. Since this ratio stays well below 100\%, it indicates that historical features contribute much more to predictive performance.
Moreover, 
Figure~\ref{motivation1}(b) shows the ratio of average gradient $\ell_2$-norms for non-historical versus historical features throughout training. The ratio below one indicates that the model relies more on historical features, making it less optimized for CS items that depend solely on non-historical features.
%
To address this, we introduce a residual connection for non-historical features, which substantially improves their updates, while incurring only a minimal increase in training cost.

To validate the third challenge in observed engagement bias against CS items, we measure the difference in average model prediction scores between CS and non-CS items, broken down by positive / negative examples for each action type. Figure~\ref{motivation1}(c) shows that the prediction scores for CS positive items are 8-14\% lower than non-CS positive. This means the ranking model tends to predict 8-14\% lower scores for a user's action on CS items than non-CS items. To address this issue, we add a regularization loss between CS item scores and non-CS item scores. This regularization improved the CS problem without any additional computational overhead.

For the last challenge in data sparsity for CS items, we propose a data-agnostic augmentation to improve model generalization, inspired by recent advances in representation learning \textit{manifold mixup}.  \cite{zhang2017mixup,verma2019manifold}.
Our approach introduces linear combinations of training embeddings as additional supervision, with the same transformation applied to their labels. This encourages the model to behave linearly between training examples, yielding a feature space that captures more diverse directions and thereby improves generalization. In particular, the augmented model learns a higher-rank feature space, reflecting more distributed representations and more effective utilization of the embedding space.
The main contributions of this paper are summarized as follows:
\begin{itemize}
\item \textbf{Comprehensive cold-start root-cause analysis in industrial RecSys}: We systematically analyze the three root causes of the multi-factored CS problem. The comprehensive analysis guided us to develop a multi-pronged approach rather than a single solution at industrial scale.

\item \textbf{Cost-efficient \& plug-and-play solutions}: The proposed solutions are cost-efficient and highly generalizable since they can augment any ranking models to enhance their cold-start capability while requiring virtually no modification to the original model. This minimizes the implementation barrier in latent-model production environments.

\item \textbf{Proven upon deployment on a leading global visual discovery platforms}: The proposed methods has been deployed with neutral cost on Pinterest's Related Pins surface \cite{liu2017related} with about 10\% increase on fresh content engagement, efficiently serving over 570 millions of users. 




    
\end{itemize}

\section{Related Works}

Early studies identified the CS problem in RecSys as a result of label sparsity and the long-tail distribution of user-item interactions \cite{wen2022distributionally, park2008long, abdollahpouri2017controlling}. Real-world RecSys benchmarks are highly imbalanced with the majority of instances offer rich historical features, a few items and users lack rich engagement data \cite{DICE, MACR, KDCRec, CausE}. Models trained on such data are susceptible to engagement bias, tending to over-recommend popular items and under-recommend those in the minority, thus amplifying CS challenges \cite{zhang2022incorporating, joachims2017unbiased}.

To address these issues, researchers have explored post-processing re-ranking \cite{RankALS, FPC, PC_Reg}, balanced training objectives \cite{ESAM, ALS+Reg, Regularized_Optimization}, sample re-weighting methods \cite{IPS-C, IPS-CN, Propensity_SVM-Rank, YangCXWBE18, UBPR}, and causal inference techniques \cite{CausE, PDA, DecRS, MACR}. While these methods have shown success in improving metrics for CS items, they are often less practical for industrial settings, as they either require access to a additional dataset for model tuning or require significant additional computational resources or extra training data~\cite{zhu2020recommendation, wei2021contrastive, yin2020learning, zhang2022incorporating}.



Cost-efficient methods, \eg, dropout on historical features, face the feature selection issue and usually leads to tradeoff between CS and non-CS items \cite{volkovs2017dropoutnet, zhao2023bootstrapping}. To mitigate historical feature sparsity issue, content-based methods are proposed to model distributions of content features \cite{wei2021contrastive,barkan2019cb2cf}. However, their impact is limited due to the gap between the semantic information and the recommendation tasks \cite{jiang2024prompt, chen2023win}, \ie, users do not always engage with items that are semantically similar to each other or to their search queries.
Knowledge distillation has been proposed to address under-prediction on CS items by transferring knowledge from a teacher trained on warm items to a student trained on cold ones \cite{zhu2020recommendation, wei2021contrastive}. However, it requires careful tuning and adds computational and infrastructure overhead due to the extra student tower.
Contrastive loss based distillation methods also require paired training data \cite{zhu2020recommendation, wei2021contrastive, zhang2022incorporating}, which are not always available in industrial setting.
Transfer learning approaches alleviate cold-item sparsity but incur extra cost by requiring an additional model \cite{barjasteh2016cold, barkan2019cb2cf, zhu2020recommendation}.

This work distinguishes itself by offering plug‑and‑play solutions that address cold‑item sparsity, limited historical feature coverage, low importance of non‑historical features, and prediction bias toward cold items, while maintaining cost efficiency.

\section{Proposed Method}

\subsection{Notation}
Throughout this section, we focus on the fundamental setting and present notation that applies to a general RecSys framework. Section \ref{experimentsatpinterest} provides details specific to the Pinterest RecSys ranking environment, including implementation and architecture.

Suppose $\mathcal{D} = {\left\lbrace (\mathbf{x}_i^h,\mathbf{x}_i^{nh} , y_i) \in \mathcal{X}_h \times \mathcal{X}_{nh}\times \mathcal{Y}\right\rbrace }_{i=1}^{n}$ is the training dataset consisting of $n$ training instances where $\mathcal{X}_h$ and $\mathcal{X}_{nh}$ represent historical and non-historical inputs, respectively, $\mathcal{Y} = {\left\lbrace0,1\right\rbrace}^m$ is the label set and  $m$ denotes the prediction tasks. Note that our formulation is for a multi-task learning and the vanilla single-task fits into our framework when $m=1$.
For ease of notation, the preliminary step of transforming raw input signals, \eg, image, video, text, and interactions, into embedding is omitted and all the training instances are vectorized, \ie, $\bx_h \in \mathbb{R}^{d_{h}} $ and  $\bx_{nh} \in \mathbb{R}^{d_{nh}} $.
Interaction module  $I(\cdot):\mathbb{R}^{d_{h}+d_{nh}}\rightarrow\mathbb{R}^{d_I}$ learns interactions across input features. Specifically, We follow the widely adopted RecSys architecture in which the historical and non-historical inputs are first passed through the
interaction module. Then the prediction module $F(\cdot):\mathbb{R}^{d_I}\rightarrow\mathbb{R}^m$, transforms the resulting embedding to $m$ prediction scores $\hat{\by} \in \mathbb{R}^m$. 

The forward pass of the mentioned model can be formalized as:
\begin{equation}\label{FF}
\begin{aligned}
   \bz&=I\left(\left[\bx^h;\bx^{nh}\right]\right), & \\
    \hat{\by}&=F(\bz),&
\end{aligned}
\end{equation}
where $[\mathbf{a};\mathbf{b}]$ denotes the concatenation of vectors $\mathbf{a}$ and $\mathbf{b}$.
For notational convenience, we define $\mathbf{\Theta}$ as the trainable parameters for $I$ and $F$. Thus, the training optimization can be defined as:
\begin{equation}\label{empericalrisk}
\begin{aligned}
   \mathbf{\Theta}^{*}  = \underset{\mathbf{\Theta}}{\arg\min } \:\:  \underset{(\bx^h, \bx^{nh},y)\sim \mathcal{D}}{\mathbb{E}} \left[ L\left(\hat{\by},\by\right)\right],
\end{aligned}
\end{equation}
where empirical risk minimization is performed using binary cross-entropy or mean squared error as $L$ \cite{guo2023embedding}.

\subsection{Non-Historical Feature Utilization} \label{secRes}
Current RecSys ranking model largely relies on historical data for prediction \cite{huang2023aligning, pan2019warm, zhu2020recommendation, zhu2021learning}. Prior works \cite{huang2023aligning, zhang2023cold, lee2023flexible, li2022inttower} have attempted to address this issue by introducing an additional embedding tower dedicated to processing non-historical data. While effective in mitigating the CS problem, the extra embedding tower substantially increases computational overhead and memory footprint. Moreover, having an extra tower will result in additional hyperparameter to tune the final output of the model which is not intuitive. 
To better understand the source of this over-reliance, we performed a gradient analysis during training. Specifically, Figure~\ref{motivation1}(b) illustrates that the ratio of average gradient $\ell_2$-norms for non-historical versus historical features is consistently bellow one, \ie, gradient for historical features is larger than that for non-historical inputs. This indicates that the model updates are dominated by historical signals, diminishing the influence of non-historical data.

\begin{figure}[t]
\centering
\includegraphics[width=1.0\linewidth]{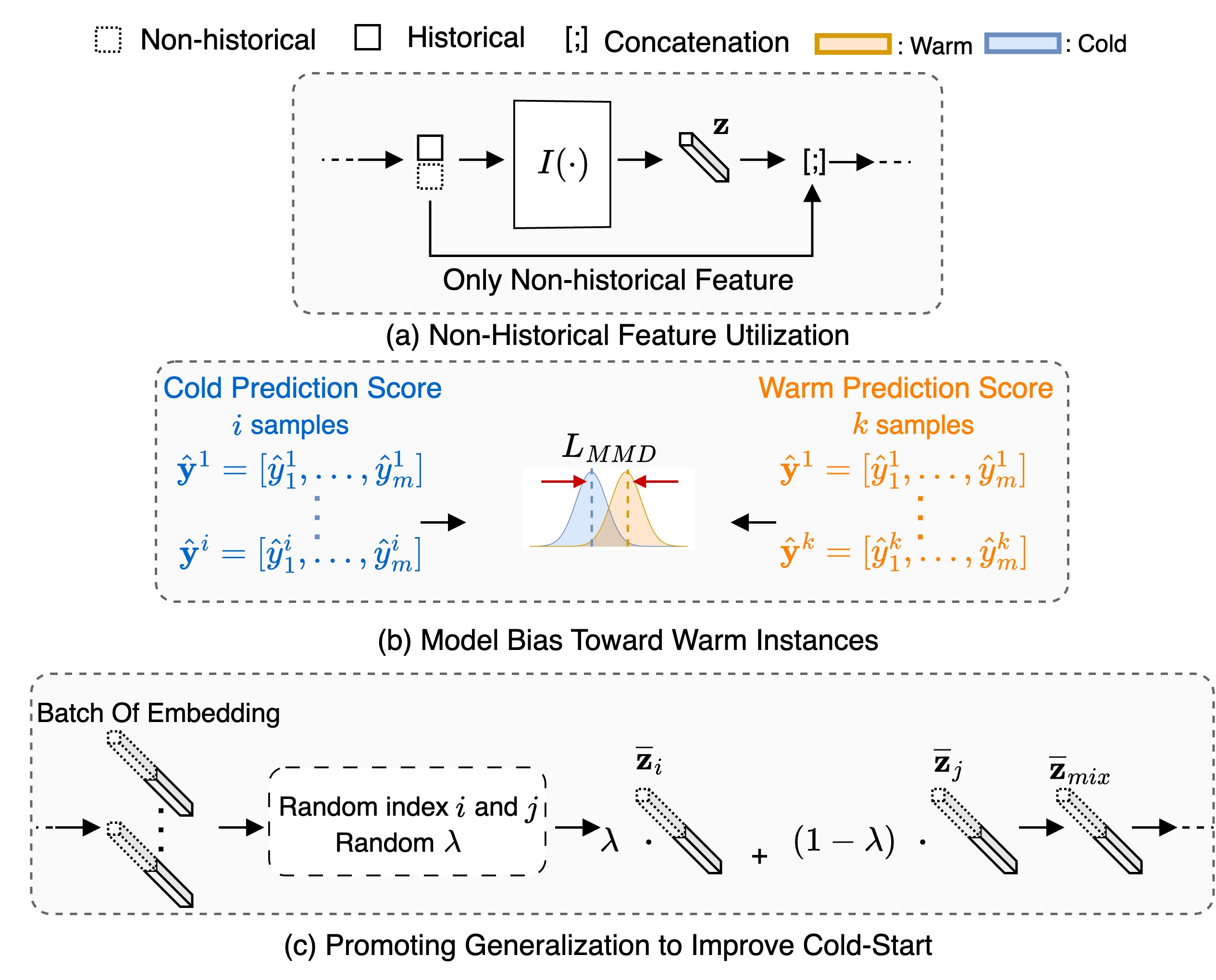}
\caption{(a) A residual connection to enhance the influence of non-historical features on model predictions. (b) A debiasing loss to reduce model bias toward warm instances, without requiring sampling techniques or computational overhead to the training. (c) Illustration of the mixup procedure: for each sample in a batch, a distinct random sample is chosen, and a new instance is generated via linear interpolation.   }\label{mainfig}

\end{figure}

Inspired by this observation, we introduce a residual path from the non-historical signals directly to the prediction module $F$. Then, the model forward path can be reformulated as:
\begin{equation}\label{res_pred}
\begin{aligned}
    \hat{\by}=F\left(\left[\bz;\bx^{nh}\right]\right).
\end{aligned}
\end{equation}
This skip connection allows non-historical features to bypass the interaction module and directly influence the model's predictions, see Figure~\ref{mainfig}a. As demonstrated in our ablation studies, this additional path from $\bx^{nh}$ to the prediction module $F$ effectively alleviates gradient issues for non-historical signals. Notably, the residual connection slightly increases the number of model parameters, \ie, less than five percent, which is substantially lower than the overhead introduced by adding an extra embedding tower, \ie, more than 28 percent. Furthermore, our approach does not require auxiliary prediction heads or complex hyperparameter tuning, making it both convenient and practical for industrial deployment.

\subsection{Model Bias Toward Warm Instances}\label{secReg}
Observations in Figure~\ref{motivation1}(c) show that when the model is naively trained on large-scale, real-world datasets, it develops a strong preference for history-rich, or "warm," instances. Specifically, let $\mathcal{D} = \mathcal{D}_{w} \cup \mathcal{D}_{c}$, where $\mathcal{D}_{w}$ and $\mathcal{D}_{c}$ denote the sets of warm and cold instances, respectively. The main pattern observed in Figure~\ref{motivation1}c can be summarized as: 
\begin{equation}\label{scorebias} 
\underset{\hat{\by} \sim \mathcal{D}_{w}}{\mathbb{E}} \left[ \hat{\by}\right] \geq \underset{\hat{\by} \sim \mathcal{D}_{c}}{\mathbb{E}} \left[ \hat{\by}\right],
\end{equation}
where $\underset{\hat{\by} \sim \mathcal{D}}{\mathbb{E}}[\hat{\by}]$ denote the expectation over model predictions drawn from the score distribution induced by inputs from $\mathcal{D}$.
Equation \ref{scorebias} illustrates that the model assigns higher scores to warm samples than cold, leading to recommendations being dominated by warm items. Previous studies have attempted to address this issue using post-processing \cite{Calibration, RankALS, rescale, FPC, PC_Reg}, balanced training \cite{ESAM, ALS+Reg, Regularized_Optimization, PC_Reg}, and sample re-weighting strategies \cite{IPS-C, IPS-CN, Propensity_SVM-Rank, YangCXWBE18, UBPR}.

However, such methods often require access to additional data or sacrifice overall performance, making them impractical for industrial use.
To overcome these obstacles, we estimate the prediction bias between cold and warm score distributions. We then penalize this bias using the Maximum Mean Discrepancy (MMD) \cite{gretton2012kernel}.: 
\begin{equation}\label{scorereg}
\begin{aligned}
 \sup_{\|F\left(I\left(\cdot\right)\right)\|_{\mathcal{H}}\leq 1}\left(\underset{\hat{\by} \sim \mathcal{D}_{w}}{\mathbb{E}} \left[ \hat{\by}\right] -\underset{\hat{\by} \sim \mathcal{D}_{c}}{\mathbb{E}} \left[ \hat{\by}\right]\right), 
\end{aligned}
\end{equation}
where $\mathcal{H}$ is a reproducing kernel Hilbert space \cite{gretton2012kernel}. Because the true ground-truth data distributions are unknown, we instead use their empirical estimation: 
\begin{equation}\label{empericalMMD}
\begin{aligned}
   L_{MMD}=\Vert\frac{1}{\vert\mathcal{D}_{w}\vert} \sum_{\left(\bx_h,\bx_{nh}\right) \in \mathcal{D}_{w}}\hat{\by} - \frac{1}{\vert\mathcal{D}_{c}\vert} \sum_{\left(\bx_h,\bx_{nh}\right) \in \mathcal{D}_{c}}\hat{\by} \Vert^2,
   \end{aligned}
\end{equation}
where $|\mathcal{D}|$ represents the cardinality of the set $\mathcal{D}$. Our proposal 1) does not add extra trainable parameters, 2) incurs negligible computational overhead during training, and 3) does not increase serving costs or latency, making it feasible for industrial deployment without requiring a specific sampling.

\subsection{Promoting Generalization to Improve Cold-Start}\label{secMix}
Prior works have generally viewed the CS problem as an outcome of the sparsity of cold instances in the training data and formulate it as a long-tail challenge \cite{wen2022distributionally, lin2024temporally, park2008long, abdollahpouri2017controlling}. A straightforward solution is to use augmentation techniques to create more cold training instances \cite{cui2015data,fawzi2016adaptive,mikolajczyk2018data}. However, designing effective augmentation methods is both costly and challenging in RecSys, and often leads to performance trade-off \cite{yang2023loam, yin2020learning, zhang2022incorporating}.

Instead, we approach the CS problem as a domain generalization challenge, where cold instances may follow a different distribution than warm samples. Domain generalization has long sought to address poor model performance under distribution shifts between training and testing data \cite{blanchard2011generalizing,li2017deeper,li2018learning}. Various remedies have been proposed, including domain-invariant feature learning \cite{muandet2013domain, ghifary2015domain, motiian2017unified, li2018domain,carlucci2019domain, wang2020learning}, meta-learning \cite{li2018learning, balaji2018metareg, dou2019domain}, data augmentation \cite{shankar2018generalizing, volpi2018generalizing, zhou2020deep}, and embedding space manipulation \cite{verma2019manifold, zhang2017mixup, li2021simple}.

We inspire from \textit{manifold mixup} \cite{verma2019manifold} to improve model generalization by training on linear interpolations of embeddings, which can be categorized on the embedding space manipulation. Specifically, the model is encouraged to explore meaningful regions of the representation space by training on linear interpolations of embeddings \cite{verma2019manifold, zhang2017mixup}. We particularly adhere to  embedding space manipulation instead of other domain generalization approaches as it does not  require  constructing of training pairs, heavy computational overhead, additional training time, or memory resources, an infeasible demand at industrial scale \cite{liu2021contrastive,xia2021self,xie2022contrastive,huang2023aligning,li2018learning, wang2020learning}.

\textit{Manifold mixup} can be integrated into the training pipeline with minimal computational overhead, as the linear interpolation operates directly on feature vectors. For each training sample undergoing \textit{manifold mixup}, the same linear interpolation is also applied to the associated labels:
\begin{equation}\label{mixup}
\begin{aligned}
   \bz_{mixed} = \lambda \cdot \bz_i + \left(1-\lambda\right)\cdot \bz_j, \\
   \by_{mixed} = \lambda \cdot \by_i + \left(1-\lambda\right)\cdot \by_j
   \end{aligned}
\end{equation}
where $i\neq j$, and $i$, $j$, are two random indices selected from the mini-batch, with $\lambda \sim \beta(\alpha,\alpha)$. This interpolation creates a novel training instance, encouraging the model to behave linearly between training examples. 
Please note that \textit{mixup} can be applied at the arbitrary layer of the model and Equation~\ref{mixup} employs \textit{mixup} at the output of interaction module for the purpose of better illustration. Please refer to \cite{verma2019manifold} for detailed explanation.

Importantly, \textit{mixup} does not rely on any specific sample selection technique; $i$ and $j$ are drawn randomly from the mini-batch. While one could enforce mixing between warm and cold samples, in practice the scarcity of cold samples reduces diversity leading to overall performance degradation. 
Additionally, while the original mixup approach uses only the mixed samples for training, we find that better performance is attained by training on both mixed and original samples. Thus, the forward pass of the model can be reformulate as: 
\begin{equation}\label{FFnew}
\begin{aligned}
    \bz               &= I\left(\left[\mathbf{x}^h; \mathbf{x}^{nh}\right]\right) & &\\
    \overline{\bz}    &= \left[\bz; \mathbf{x}^{nh}\right] \\
    \overline{\bz}_{mix},\, \by_{mix}& = \mathrm{mixup}(\overline{\bz}, \by) & \\
    \hat{\by},\, \hat{\by}_{mix}        &= F(\overline{\bz}),\, F(\overline{\bz}_{mix})        &    &
\end{aligned}
\end{equation}
As a result, the training objective should be computed for both $\left(\hat{\by},\by\right)$ and $\left(\hat{\by}_{mix},\by_{mix}\right)$:
\begin{equation}\label{FFnew}
\begin{aligned}
     L\left(\hat{\by},\by\right) + \lambda_1 \, L\left(\hat{\by}_{mix},\by_{mix}\right)
\end{aligned}
\end{equation}
where $\lambda_1$ is a balancing factor that controls the extent to which {mixup} instances are incorporated during training.
\subsection{Overall Training Objective}
Our proposal includes includes two additional training objectives to the original training framework. Specifically, the overall training objective can be defined as:
\begin{equation}\label{res_pred}
\begin{aligned}
    L_{final}= L\left(\hat{\by},\by\right) + \lambda_1 \, L\left(\hat{\by}_{mix},\by_{mix}\right) + \lambda_2 \,L_{MMD}\left(\hat{\by},\by\right)
\end{aligned}
\end{equation}
where $\lambda_1$ and $\lambda_2$ represent the weights for the loss computed from the mixed samples and MMD loss, respectively.

\begin{figure}[t]
  \centering
  \includegraphics[width=0.65\linewidth]{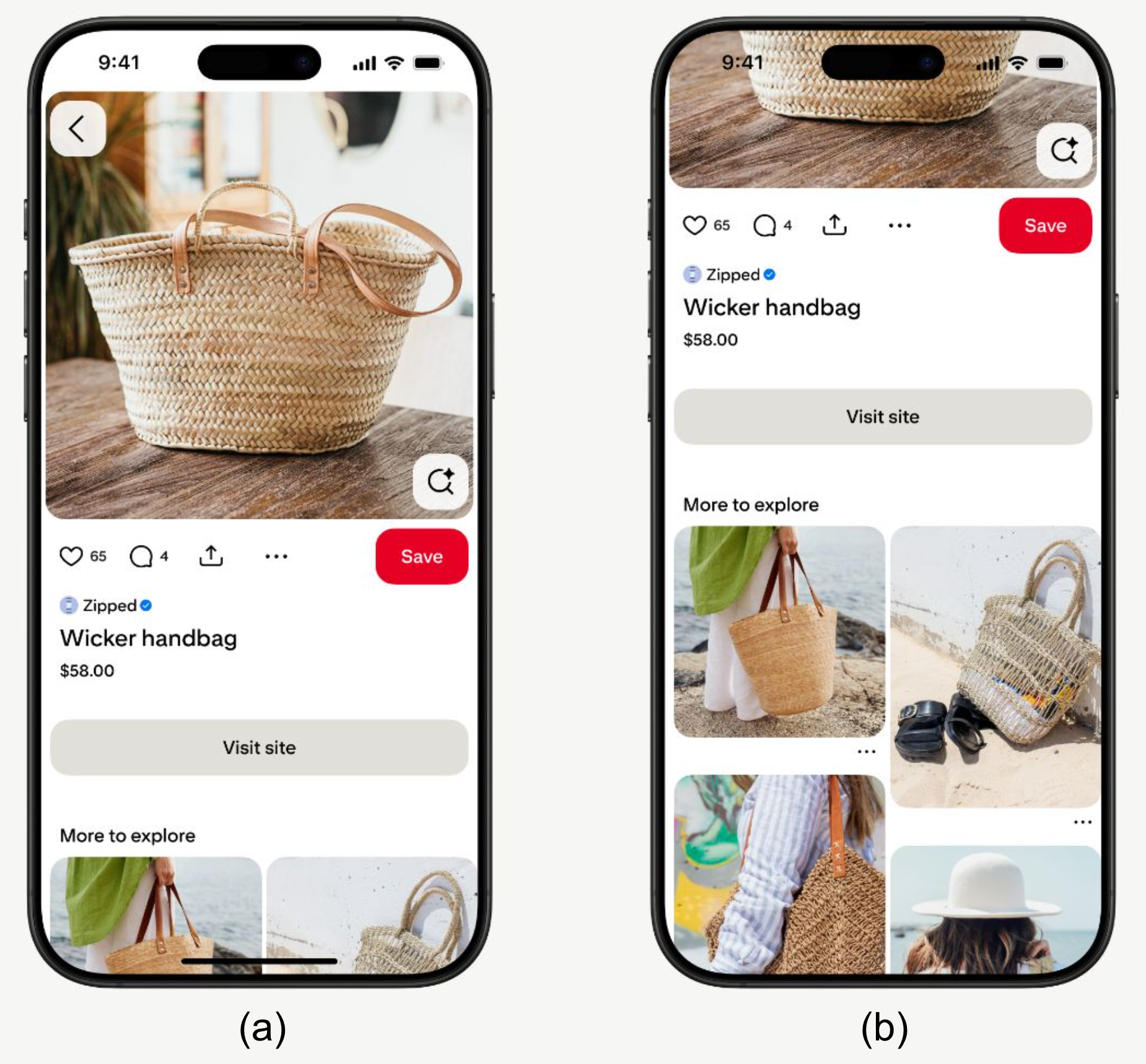}
  \caption{Related Pins Surface on Pinterest. (a) The Pin, \ie, query Pin, user just grid-clicked on from an upper-stream surface. (b) Grid view of
recommended Pins. User name and profile are masked out for privacy protection.}\label{relatedpinsurface}
  \Description{Related Pins Surface on Pinterest (User name and
profile are masked out for privacy protection).}
\end{figure}

\section{Experiments} \label{experimentsatpinterest}


\subsection{Experimental Settings}
We evaluated the proposed methods on Pinterest Related Pins surface \cite{liu2017related}, a search-like surface which serves around 50\% of the traffic on the entire platform.
Users enter the Related Pins surface when they grid-click a Pin on any other upper-stream surface (such as Homefeed or Search). There are two main parts of the Related Pins surface: (1) A closer view of the Pin (query Pin) they just grid-clicked on from an upper-stream surface (Figure~\ref{relatedpinsurface}a), and (2) When users scroll down, they will see a grid view of recommended Pins under title ``More to explore" which are related to the query Pin above (Figure~\ref{relatedpinsurface}b). From the grid view, users can \textit{grid-click} on a Pin to take a closer view of it (e.g., read the title and descriptions) and \textit{click} to visit the link associated with the item. Users can also \textit{save} a Pin to their collections on own profile.





\begin{figure}[t]
  \centering
  \includegraphics[width=0.7\linewidth]{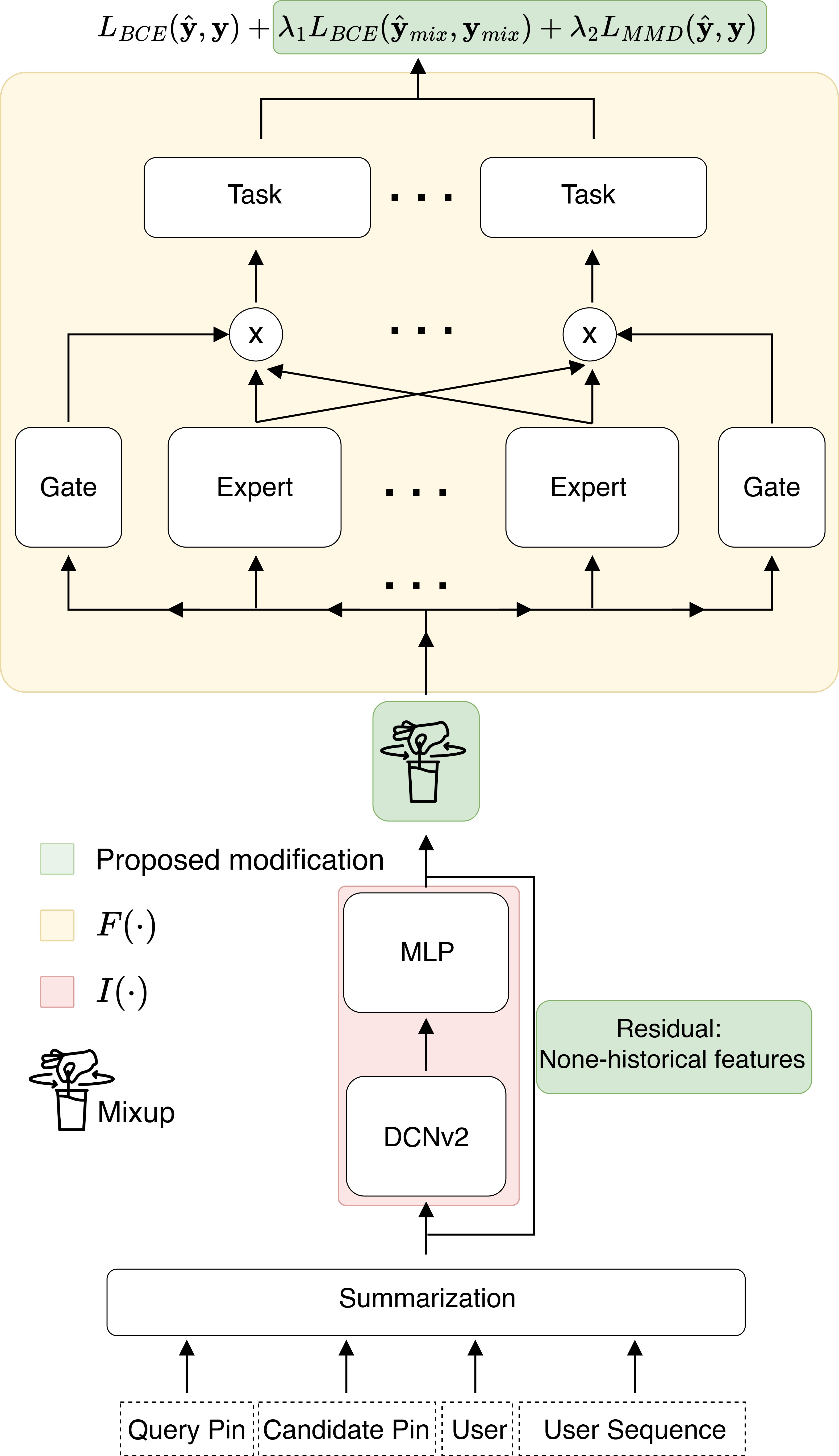}
  \caption{Multi-task Ranking Model Architecture in Pinterest Related Pins Recommender System}\label{relatedpinarch}
  \Description{Pinterest Related Pins Multi-task Ranking Model Architecture}
\end{figure}

The main architecture of the ranking model in Pinterest Related Pins recommender system is designed to optimize for multi-task learning across related tasks, including \textit{grid-click}, \textit{save}, and \textit{click} etc, as showin in Figure~\ref{relatedpinarch}. The ranking model's inputs include query Pin features, candidate Pin features, user features, along with a transformer-based user sequence embedding that captures the user’s historical events. The summarization module groups related features and learns certain feature crossings. The outputs from the summarization layer are processed by the DCNv2 module \cite{wang2021dcn} to learn higher-order feature interactions. The resulting outputs are then concatenated into a single feature vector and reduced dimension through MLP. Finally, the aggregated embedding is passed through an MMoE module \cite{ma2018modeling} to produce the final prediction scores for each task. 

The label space for each task is binary. Based on Equation (\ref{res_pred}), the final training objective is the weighted sum of the binary cross entropy loss from the original labels, the binary cross entropy loss computed from the mixed samples, and the introduced MMD loss: 
\begin{equation}
\label{bce}
    L_{final}= L_{BCE}(\hat{\by},\by) + \lambda_1 L_{BCE}(\hat{\by}_{mix},\by_{mix}) + \lambda_2 L_{MMD}(\hat{\by},\by),
\end{equation}
The final output score for each candidate Pin is computed as:
\begin{equation}
    s = \hat{\by} \cdot \boldsymbol{u}
    \label{final_score}
\end{equation}
where $\cdot$ denotes the element-wise multiplication, and $\boldsymbol{u}$ represents the utility weight vector of all the actions (\eg, grid-click, save, click) representing the
importance of the task which is defined based on business need.



\subsection{Offline Experiments}
\subsubsection{\textbf{Datasets \& Model Training}}
We train the multi-task learning ranking model of Pinterest Related Pins surface \cite{liu2017related}  illustrated in Figure \ref{relatedpinarch} with Equation (\ref{bce}) as training objective and use Equation (\ref{final_score}) to compute the final prediction score for ranking. We use 27 days of data for training, the last day
of training data for calibration and the following three days
of data for evaluation. We empirically tuned $\lambda_1$ and $\lambda_2$ to 0.2 and 0.1, respectively. The model is trained for one epoch.

\subsubsection{\textbf{Compared Methods}} 
We compare our approach against the following approaches: (1) Baseline: the original model architecture without the proposed components (see Figure~\ref{relatedpinarch}); (2) Feature Dropout \cite{volkovs2017dropoutnet}, which randomly drops historical features; and (3) ALDI \cite{huang2023aligning}, a recent method that builds a separate tower using only non-historical features and aligns it through distillation for cold-start (CS) recommendation.
For clarity, we refer to our proposed components as follows: the residual connection for leveraging non-historical features (Section~\ref{secRes}) as \textbf{Residual}; the MMD-based loss for debiasing CS items (Section~\ref{secReg}) as \textbf{ScoreReg}; and the \textit{manifold mixup} inspired augmentation for promoting generalization (Section~\ref{secMix}) as \textbf{Mixup}.

\subsubsection{\textbf{Metrics Improvement}}
We used \textit{Hits@3} to evaluate the overall ranking performance based on the final ranking score defined in Equation (\ref{final_score}). For example, we increment \textit{Hits@3(save)} if the user saved any of the top three recommendations. At Pinterest, the business goal on CS problem is improving users' engagement, \ie, grid-click and save, on fresh Pins that are created within the recent 28 days. Therefore, we use \textit{Hits@3} on fresh Pins < 28 days as offline CS metric.

\begin{table}
\caption{Offline \textit{Hit@3} lift (\%) on fresh Pins < 28 day and all Pins and model parameter increase (\%).}
  \label{offline}
  \addtolength{\tabcolsep}{3pt}
\resizebox{1.0\linewidth}{!}{
\begin{tabular}{c|c|cc|cc}
\toprule
                        &                                   & \multicolumn{2}{c|}{Fresh Pins} & \multicolumn{2}{c}{All Pins}    \\ \cline{3-6} 
\multirow{-2}{*}{Model} & \multirow{-2}{*}{\begin{tabular}[c]{@{}c@{}}Params \\ increase \end{tabular}} & Grid-click     & Save           & Grid-click     & Save           \\ 
\midrule
Dropout & 0 & +0.28 & -1.02 & -0.13 & \textbf{-0.28} \\
ALDI & +28.89 &\textbf{+0.63} & -0.19 & \textbf{+0.30} & +0.02\\
\midrule
Residual (1) & +4.96 &\textbf{+1.32}  & \textbf{+1.61} & +0.07 & -0.09  \\
ScoreReg (2) & 0 &\textbf{+1.59} & \textbf{+1.75}  & +0.01  & +0.12 \\
Mixup (3)    & 0 & \textbf{+2.03}   & \textbf{+1.42}   & -0.27 & -0.12 \\
(1) \& (2) & +4.96 & \textbf{+2.00} & \textbf{+2.15} & +0.03 & +0.05 \\
(1) \& (3) &  +4.96 & \textbf{+2.56} & \textbf{+3.24} & +0.02& +0.06\\
(2) \& (3) & 0 & \textbf{+3.16} & \textbf{+2.77} & -0.13 & -0.21\\ \midrule
\rowcolor[HTML]{C0C0C0} (1) \& (2) \& (3)  & +4.96 & \textbf{+3.20} & \textbf{+3.40} & -0.04 & -0.13 \\
 \bottomrule
\end{tabular}}
\begin{tablenotes}
    \item Note: Bold numbers are stats sig with p<0.05. Non-bolded numbers (e.g. Most of All Pins metrics) mean neutral results.
\end{tablenotes}
\end{table}

 \begin{figure*}[t]
  \centering
  \includegraphics[width=1.0\linewidth]{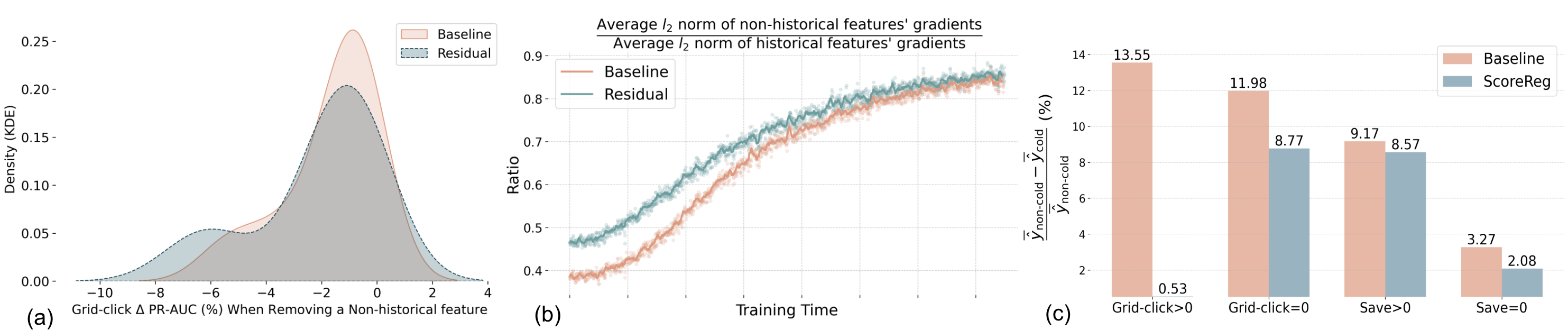}
  \caption{(a) Grid-click $\Delta$ PR-AUC distribution of removing any of the non-historical features in the baseline model v.s. Residual. The left shifted distribution in Residual indicates larger PR-AUC drop when removing some non-historical features in the proposed residual connection technique, reflecting improved feature importance of those non-historical features in the Residual model. (b) The ratio of the average $l_2$ norm of the gradients of non-historical features to that of historical features during training. Residual increases this ratio consistently, indicating that the model with residual connections on non-historical features relies more on those features during training. (c) Model prediction score gap ratio between non-cold start Pins and cold-start Pins. ScoreReg narrows the prediction gaps for both positive labels and negative labels on grid-click and save.}
  \label{score_gap}
\end{figure*}

 \begin{figure}[t]
  \centering
  \includegraphics[width=1\linewidth]{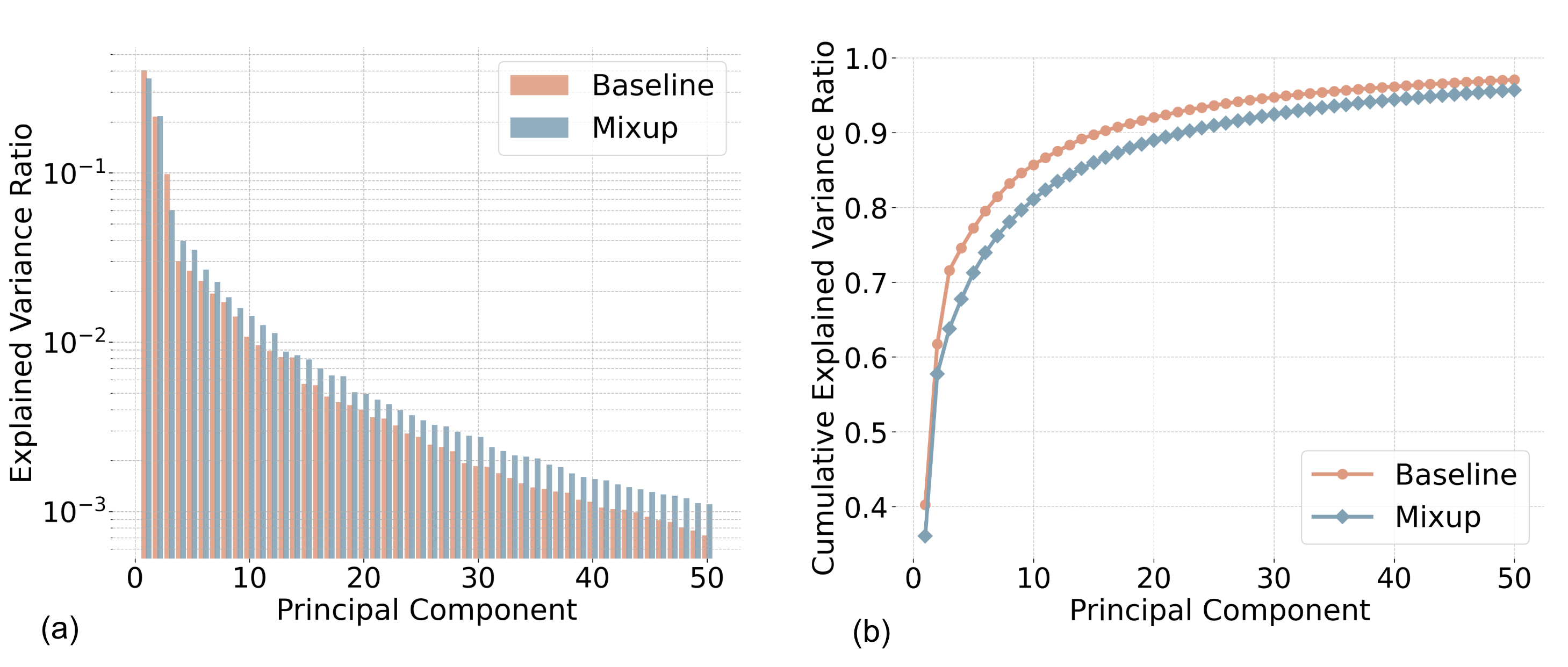}
  \caption{Embedding Space Utilization Analysis (Mixup) }
  \label{pca}
\end{figure}

Table \ref{offline} shows the offline \textit{Hits@3} lift of all the compared methods and the proposed approaches on fresh Pins and all Pins.
The proposed three techniques achieved significant improvement on fresh Pins \textit{Hits@3(grid-click)} and fresh Pins \textit{Hits@3(save)} while maintaining nearly neutral impact on overall metrics. 
We have the following observations: (1) Among all the single methods, ScoreReg achieved the highest lift of 1.75\% on fresh Pins \textit{Hits@3(save)} and Mixup achieved the best lift of 2.03\% on fresh Pins \textit{Hits@3(grid-click)}. In general, all the proposed three methods achieved better metrics gain on fresh Pins than any of the compared methods. (2) Combining the proposed three methods together accumulate the gains to a large degree, \ie, 3.2\% on fresh Pins \textit{Hits@3(grid-click)} (80\% of the additive gain) and 3.40\% on fresh Pins \textit{Hits@3(save)} (71\% of the additive gain). (3) ScoreReg and Mixup are cost neutral since they just add new losses during training without changing the model architecture and serving. Residual increases the model parameters by 4.96\% from adding a MLP layer to lower the dimensions for non-historical features before adding to the final MLP layer. (4) Dropout on historical features is cost neutral but it has significant metrics tradeoff on save for both fresh Pins and all Pins. It is hard to select which (combination of) historical features to dropout. ALDI achieved +0.63\% lift on fresh Pins grid-click and +0.3\% lift on all Pins grid-click, but its lift on fresh Pins is lower than any of the proposed methods and most importantly, it increases the model's parameters by 28.89\% from the additional content features based student tower, incurring non-trivial infra cost for both model training and serving.


\subsubsection{\textbf{Feature Importance \& Gradient Analysis (Residual)}}  
The proposed residual connection mitigates the negative impact of lack of historical features for CS items and thus contributed to fresh Pins engagement metrics gain. First, it improves the feature importance of non-historical features. Figure \ref{score_gap}(a) plots the $\Delta$PR-AUC distribution when randomly removing a historical or non-historical feature in both the baseline and Residual models during inference.
The more negative the $\Delta$PR-AUC is, the more important the feature is to the model. Figure~\ref{score_gap}(a) illustrates a left shifted distribution (more negative $\Delta$ PR-AUC) of non-historical features in the Residual model compared with the baseline model. This left shift indicates that a non-negligible number of non-historical features have their feature importance increased with the proposed residual connection technique.
Second, residual connection improves model's reliance on non-historical features during training. Figure \ref{score_gap}(b) depicts the ratio of the average $l_2$ norm of the gradients of non-historical features to that of historical features during training. Residual increases this ratio consistently, indicating that the model with residual connections on non-historical features relies more on those features during training.

\subsubsection{\textbf{Model Prediction Score Debiasing Analysis (ScoreReg)}} 
Figure \ref{score_gap}(c) plots the difference in average model prediction scores between CS and non-CS items, broken down by
positive / negative instances.
The baseline model over-predicts non-CS Pins while under-predicts CS Pins. Notably, ScoreReg narrows the prediction gaps for both positive labels and negative labels on grid-click and save, for example, it reduces the gap on predicting positive grid-clicks from  13.65\% to 0.53\%. This demonstrates that ScoreReg learns to be less biased towards CS Pins.

\subsubsection{\textbf{Embedding Space Utilization Analysis (Mixup)}}

To understand why baseline model fails to generalize to CS instances, we conducted extensive preliminary investigations and identified model's over-reliance on historical features (illustrated in Figure \ref{motivation1}a and Figure \ref{motivation1}b). Naively trained models tend to take shortcuts, heavily exploiting the dominant historical signals in the training set. Verma \etal \cite{verma2019manifold} has shown that enforcing linear behavior through mixup training flattens hidden representations and reduces the number of directions with significant variance. To quantify this effect in our setting, we analyze the effective information in the model's feature space using Principal Component Analysis (PCA). Specifically, Figure~\ref{pca}(a) visualizes the explained variance ratio across principal components and Figure~\ref{pca}(b) illustrates the cumulative explained variance ratio. 
The results clearly show that the mixup model’s feature space is less constrained and high-rank than that of the baseline model. This observation aligns with the improved performance of the mixup-trained model and with the theoretical analysis of Gunasekar \etal \cite{gunasekar2017implicit} which suggests that low-rank feature spaces hinder generalization by memorizing trivial patterns.
This supports the view that mixup mitigates shortcut reliance by promoting smoother embeddings that generalize better to CS instances.

\subsection{Online A/B Test}
\begin{table}
\centering
\caption{Online metrics volume lift (\%) on fresh Pins < 28 day and all Pins}
  \label{online}
\addtolength{\tabcolsep}{8pt}
\resizebox{1.0\linewidth}{!}{
\begin{tabular}{l|cc|cc}
\toprule
\multirow{2}{*}{Model} & \multicolumn{2}{c|}{Fresh Pins} & \multicolumn{2}{c}{All Pins} 
\\ 
\cmidrule{2-5}
 & Grid-click & Save & Grid-click&  Save \\
\midrule
Residual & \textbf{+4.48}  & \textbf{+5.98} & \textbf{+0.36} & -0.23  \\
ScoreReg& \textbf{+4.07} & \textbf{+7.88} & \textbf{+0.35} & -0.18 \\

Mixup & \textbf{+3.19}  & \textbf{+2.45}  & -0.02 & +0.14 \\ \midrule
\rowcolor[HTML]{C0C0C0} All & \textbf{+9.68} & \textbf{+9.62} & \textbf{+0.26} & -0.32 \\
 \bottomrule
\end{tabular}}
\begin{tablenotes}
    \item Note: Bold numbers are stats sig with p<0.05. Non-bolded numbers (e.g. All Pins Save) mean neutral results.
\end{tablenotes}
\end{table}

\begin{table}
\centering
\caption{Online metrics lift (\%) on number of unique fresh Pins with engagement}
  \label{online2}
\addtolength{\tabcolsep}{12pt}
\resizebox{1.0\linewidth}{!}{
\begin{tabular}{c|c|c}
\toprule
Model & with 10+ Impressions & with 10+ Saves\\
\midrule
Residual & \textbf{+1.49} & \textbf{+7.82}\\
ScoreReg & \textbf{+3.34} & \textbf{+5.12}
\\
Mixup & -0.73 & \textbf{+3.95} \\
\bottomrule
\end{tabular}}
\begin{tablenotes}
    \item Note: Bold numbers are stats sig with p<0.05. Non-bolded numbers mean neutral results.
\end{tablenotes}
\end{table}

\subsubsection{\textbf{Setup}} We conduct large-scale online A/B experiments on Pinterest Related Pins
surface \cite{liu2017related}. Given the negative offline metrics of historical features dropout and the significant increase of model size and infra cost that ALDI incurs and its weaker metrics gain on fresh Pins, we deploy the proposed three methods for online A/B tests. The number of users in each group is $\sim$ 12 million. 

\subsubsection{\textbf{Metrics Improvement on fresh Pins and all Pins}} Aligned with the offline metrics in Table \ref{offline}, the proposed three methods show stats sig metrics gain of 2.5\% to 6\% on fresh Pins (Table \ref{online}). Residual achieved the highest gain of 4.48\% on fresh Pins grid-click and ScoreReg achieved the best gain of 7.88\% on fresh Pins save. Combining all the three methods show 9.68\% aggregated gain on fresh Pins grid-clicks and 9.62\% aggregated gain on fresh Pins saves. 

Table \ref{online2} further shows that both Residual and ScoreReg help signficantly increase the number of unique fresh Pins with 10+ impressions by 1.49\% and 3.34\%, respectively. All the three methods help increase the number of unique fresh Pins with 10+ saves by 4\% to 8\%.

\subsubsection{\textbf{Metrics Improvement on cold-start users}} At Pinterest, CS users are defined as users who are active less than a certain number of days in the last 28 days. One of the most important metrics to evaluate user activeness at Pinterest is total successful sessions, which refers to the number of sessions that include at least one positive action (\eg, grid-click, save, click \etc). Accordingly, a grid-click session means there is at least a grid-click action in a session and a save session means there is at least a save action in a session. Table \ref{cold-users} shows that ScoreReg achieved 0.23\% stats sig total successful sessions lift, conributed by 0.29\% stats sig grid-click sessions gain and 0.28\% stats sig save sessions gain. Mixup achieved 0.08\% stats sig total successful sessions gain and 0.1\% stats sig grid-click sessions gain. In addition, ScoreReg and Mixup also achieved positive save volume gain for CS users on the two upper stream surfaces Homefeed and Search, which indicating a positive feedback loop of CS users being more active helps enrich their user histories in their user sequences, the most important user features of the ranking models on Homefeed and Search, and therefore enhance their ranking performance on CS users.

\begin{table}[t]
\caption{Online metrics (volume) lift (\%) for cold-start users}
  \label{cold-users}
\addtolength{\tabcolsep}{4pt}
\resizebox{1.0\linewidth}{!}{
\begin{tabular}{c|c|c|c|c}
\toprule
Metrics & Residual & ScoreReg & Mixup & All\\
\midrule
Successful sessions & +0.12 & \textbf{+0.23} & \textbf{+0.08} & \textbf{+0.13}\\
Grid-click sessions & +0.18 & \textbf{+0.29} & \textbf{+0.10} & \textbf{+0.14}\\
Save sessions & +0.04 & \textbf{+0.28} & +0.06 & \textbf{+0.21}\\
Save (Homefeed) & +0.11 & +0.50 & \textbf{+0.77} & \textbf{+0.41} \\
Save (Search) & +0.50 & \textbf{+0.84} & \textbf{+0.60} & +0.22\\
 \bottomrule
\end{tabular}}
\begin{tablenotes}
    \item Note: Bold numbers are stats sig with p<0.05.
\end{tablenotes}
\end{table}

\subsubsection{\textbf{Metrics Improvement on Lowered Ranked Pins (Mixup)}} Table \ref{tail} shows that Mixup improved both the grid-click and save metrics on lower ranked Pins significantly. This is because Manifold mixup smooths the representation space by better utilizing the principal components (as illustrated in Figure \ref{pca}). By enforcing approximate linearity between interpolated hidden states, it enhances generalization to rare feature configurations and mitigates memorization of head-dominant patterns, thereby improving rank sensitivity for long‑tail items and reducing position bias. As users engage with more content on the recommendation list per request, more positive actions are added to their user sequences and benefit the recommendation performance on the other two surfaces Homefeed and Search (illustrated in Table \ref{cold-users}). 

\textbf{The three proposed methods have been launched into the production ranking model on Pinterest Related Pins surface \cite{liu2017related} and are efficiently serving over 570 millions of users without incurring additional computational costs.}



\begin{table}[t]
\small
\caption{Online metrics (volume) lift (\%) by position (Mixup)}
  \label{tail}
\addtolength{\tabcolsep}{3pt}
\resizebox{1.0\linewidth}{!}{
\begin{tabular}{c|ccccc}
\toprule
Action & top 10 & 10-25 & 25-50 & 50-100 & beyond 100\\
\midrule
Grid-click & -0.10 & +0.11 & \textbf{+0.27} & \textbf{+0.45} & \textbf{+0.63}\\
Save & -0.06 & \textbf{+0.31} & 
\textbf{+0.55} & \textbf{+0.92} & \textbf{+1.26} \\

\bottomrule
\end{tabular}}
\begin{tablenotes}
     \item Note: Bold numbers are stats sig with p<0.05. Non-bolded numbers mean neutral results.
\end{tablenotes}
\end{table}

\section{Conclusion}
In this work, we presented a cost-efficient framework for alleviating the cold-start problem in large-scale recommender systems at Pinterest. The proposed method rebalances the model’s dependence on historical and non-historical signals through a residual connection for non-historical features and a lightweight regularization term applied during training. Furthermore, by employing an embedding-based mixup augmentation strategy, our approach improves generalization to cold items without degrading performance on warm ones or increasing serving cost. Empirical results on Pinterest's production-scale online A/B tests demonstrate consistent improvements in ranking quality and substantial reductions in cold-start bias with minimal computational overhead. These results highlight a practical path toward more balanced, generalizable, and deployable recommendation models for real-world applications.


\bibliographystyle{ACM-Reference-Format}
\bibliography{sample-base}









\end{document}